\documentclass[prb,twocolumn,showpacs,amsmath,amssymb,superscriptaddress]{revtex4}

\usepackage{graphicx}% Include figure files
\usepackage{amssymb,amsmath}
\usepackage{dcolumn}% Align table columns on decimal point
\usepackage{bm}% bold math
\usepackage{mathrsfs}% bold math
\usepackage{color}

\begin{document}

\title{Parametric Inversion of Spin Currents in Semiconductor Microcavities}

\author{H. Flayac}
\affiliation{Institut Pascal, PHOTON-N2, Clermont Universit\'{e}, Blaise Pascal University, CNRS, 24 avenue des Landais, 63177 Aubi\`{e}re Cedex, France.}
\author{D. D. Solnyshkov}
\affiliation{Institut Pascal, PHOTON-N2, Clermont Universit\'{e}, Blaise Pascal University, CNRS, 24 avenue des Landais, 63177 Aubi\`{e}re Cedex, France.}
\author{G. Malpuech}
\affiliation{Institut Pascal, PHOTON-N2, Clermont Universit\'{e}, Blaise Pascal University, CNRS, 24 avenue des Landais, 63177 Aubi\`{e}re Cedex, France.}
\author{I. A. Shelykh}
\affiliation{Science Institute, University of Iceland, Dunhagi-3, IS-107, Reykjavik, Iceland}
\affiliation{Division of Physics and Applied Physics, Nanyang Technological University 637371, Singapore.}

\begin{abstract}
The optical spin-Hall effect results in the formation of an antisymmetric real space polarization pattern forming spin currents. In this paper, we show that the exciton-polariton parametric scattering allows us to reverse the sign of these currents. We describe the pulsed resonant excitation of a strongly coupled microcavity with a linearly polarized pump at normal incidence. The energy of the pulse is set to be close to the inflexion point of the polariton dispersion and the focusing in real space populates the reciprocal space on a ring. For pumping powers below the parametric scattering threshold, the propagation of the injected polaritons in the effective magnetic field induced by the TE and TM splitting produce the normal optical spin-Hall effect. Keeping the same input polarization but increasing the pump intensity, the parametric scattering towards an idler and a signal state is triggered on the whole elastic circle. The injected particles are scattered toward these states while propagating radially all over the plane, gaining a cross linear polarisation with respect to the pump during the nonlinear process. Eventually, the propagation of the polaritons in the effective field results in the optical spin Hall-effect, but this time with inverted polarization domains.

\end{abstract}
\maketitle

\section{Introduction}
In the domain of the mesoscopic physics,
spintronics\cite{Spintronics} is currently one of the most promising
areas. The main idea of this discipline is to control the spin of
individual carriers, based on their quantum
properties, which could have a huge impact on future
information technologies. Although currently
spintronics rely on giant magneto-resistance effect in metals only,
there are good perspectives that in future
devices, which now still remain at the stage of the theoretical
modeling, other effects will find their way to practical implementations. One of
the most serious obstacles resides in the dramatic role played by the processes of spin
relaxation.

In this context, it was proposed that the optical counterpart of
spintronics, namely \emph{spin-optronics}\cite{Spinoptronics} could
represent a valuable alternative, since the corresponding characteristic decoherence
times are orders of magnitude longer than
those of electrons and holes\cite{PolDev}. The entities
under study in spinoptronics are
exciton-polaritons\cite{Microcavities} which are the elementary
excitations of semiconductor microcavities within the strong
coupling regime. Being a mixture of quantum-well excitons and cavity
photons, they possess numerous peculiar properties distinguishing
them from other quasi-particles in mesoscopic systems. They inherit
a very light effective mass from their photonic component which
allows their ballistic propagation at large velocities
\cite{Wertz,SpinBullets}. Their excitonic part allows them to
interact efficiently with each other giving birth to strong
nonlinear phenomena such as the bistability\cite{Bistability}, the
optical parametric oscillations \cite{OPO} or the formation of an
interacting quantum fluid of light\cite{QuantumFluid} and its
topological excitations\cite{LagoudakisV,LagoudakisHV,AmoOS,AmoOHS}.

Importantly, from the point of view of their spin structure,
polaritons can be considered as a two-level system, analogous to
electrons\cite{ReviewSpin}. The two allowed spin projections
$\pm1$ correspond to the two opposite circular polarizations of the
counterpart photons. As for any two-level system, one can introduce
the concept of the pseudospin vector $\mathbf{S}$ for the
description of the polarization dynamics of polaritons. In full
analogy with the case of electrons in the context of spintronics $\mathbf{S}$ undergoes a precession caused by effective
magnetic fields, arising from intrinsic or extrinsic polarization splittings\cite{ReviewSpin}.

Most of the recent developments in spin-optronics and prospects for
its future applications\cite{PolDev} are based on the
spectacular progresses of the last decades in the engineering of nanoscale
systems and experimental investigation of their optical and
transport properties, which revealed remarkable novel spin and light
polarization effects\cite{ReviewSpin}. Among them is the optical
analog of the spin-Hall effect, proposed in 2005\cite{OSHE},
later on observed experimentally several
times\cite{OSHENPhys,OSHEOpt,AOSHE,Kammann} and recently
re-investigated theoretically\cite{NOSHE}. The spin-Hall effect (SHE)
for electrons consists in the generation of pure spin currents
perpendicular to the electric current in 2D electron
systems\cite{SHI1}. There exist two variants of this effect: the
extrinsic SHE\cite{SHI2} provided by the spin- dependent Mott
scattering of propagating electrons on impurities and the intrinsic
SHE\cite{SHI3} generated by the spin-orbit interaction (SOI) of the
Rashba type, which results in the appearance of the $k$-dependent
effective magnetic field rotating the spins of the propagating electrons.
The optical spin-Hall Effect (OSHE) is analogical to the intrinsic
SHE. The role of Rashba SOI in this case is played by TE-TM
splitting of the polariton mode\cite{Berry} that however doesn't break the time reversal symmetry due to its peculiar wavevector dependence.

In this paper we show that the spin currents created by the OSHE can be fully inverted under proper excitation of the polaritonic states. This phenomenon occurs when the parametric scattering of polaritons is triggered over an elastic circle at the magic angle in reciprocal space. The final signal and idler states gain a linear polarization that is rotated by $\pi/2$ with respect to that of the pump state subsequently inverting the spin domains over the whole cavity plane.

\section{Optical Spin-Hall effect}\label{SecI}

\begin{figure}[ht]
\includegraphics[width=0.95\linewidth]{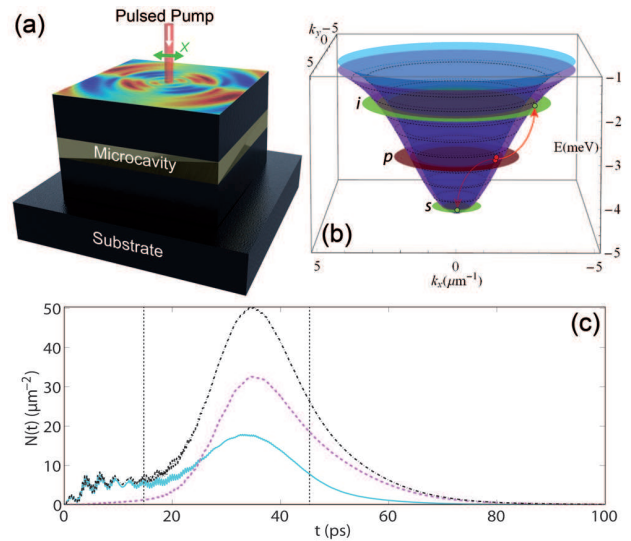}
\caption{(Color online) (a) Scheme of the microcavity illustrating
the pumping geometry. The $X$-polarized (green double arrow) injected
polaritons propagate radially outward form the narrow/pulsed pump spot and
form circular polarization domains (colormap) [see also Fig.\ref{fig3}]. (b) TE (purple surface) and TM (blue
surface) lower polariton branches (their energy splitting has been
emphasized for clarity). The
red disk illustrates the pump ($p$) excitation close to the
inflexions of the branches [see Eq.(\ref{Conservation})] that are
both excited due to the $X$-polarized pump and the smallness of the
spot in real space. The green disks mark the
signal ($s$) and idler ($i$) states appearing when the OPO is
triggered. The red arrows point an example of OPO process occurring
along $x$-direction: Two pump polaritons from one branch are
scattered toward the signal and idler states of the other branch. (c) Space-integrated densities of the $X$
(solid blue line) $Y$-polarized (dashed pink line) particles and the
sum of the two (dashed-dotted black line) revealing the
polarization inversion occuring after 25 ps and reaching
its maximum at 35 ps. The vertical dashed lines mark the times at
which the snapshots of Fig.\ref{fig3} where taken.} \label{fig1}
\end{figure}

It is well known that due to the long-range exchange interaction
between an electron and a hole, for excitons having non-zero
in-plane wavevectors, the states with dipole moments oriented along
and perpendicular to the wavevector are slightly different in
energy\cite{Maialle}. In microcavities, this splitting is amplified
due to the exciton coupling with the cavity mode\cite{Panzarini1999}
and can reach values of up to 1 meV. The TE-TM splitting results in
the appearance of a $k$-dependent effective magnetic field
$\mathbf{H}_{lt}$ provoking the rotation of polariton pseudospin $\mathbf{S}=(S_x,S_y,S_z)^T$. It
is oriented in the plane of the microcavity and makes a double angle
with respect to the wavevector:
\begin{eqnarray}
\label{HLT}
{\mathbf{H}_{lt}}\left( \mathbf{k} \right) &=& {\Delta _{lt}}\left( \mathbf{k} \right){\left( {\cos 2 \phi ,\sin 2 \phi } \right)^T}\\
\label{DeltaLT}
{\Delta _{lt}}\left( \mathbf{k} \right) &=& {E_{t}}\left( \mathbf{k} \right) - {E_{l}}\left( \mathbf{k} \right)
\end{eqnarray}
where $\phi$ is the polar angle. We remind that the $S_x$ and $S_y$ components correspond to linear polarization of the polariton emission while the $S_z$ component stands for the circularly polarized states. We measure $\mathbf{H}$ in energy units and $E_{t}$ and $E_{l}$
are the dispersion relations of the TE and TM polarized polaritons
[see Eq.(\ref{Displt})]. This orientation is imposed by the symmetry
of the TE and TM states over an elastic circle. For example, along
the $x$ direction $X$-polarized polaritons are TM while they are TE
along the $y$ direction and reciprocally for $Y$-polarized
particles. For particles propagating without scattering with a given
value of $\mathbf{k}$ and in the absence of spin relaxation, the dynamics of $\mathbf{S}$ is
governed by a simple vectorial precession equation :
\begin{equation}\label{SEq}
\frac{{\partial \mathbf{S}\left( t \right)}}{{\partial t}} = \frac{{{\mathbf{H}_{lt}}\left( \mathbf{k} \right) \times \mathbf{S}\left( t \right)}}{\hbar }
\end{equation}
corresponding to an undamped Landau-Lifshitz equation. The redistribution of the polaritons in the reciprocal space,
provoked e.g. by impurity scattering, leads to a change in the
direction of the rotation of pseudospin and can result in the formation
of polariton spin currents in the plane of the microcavity.  As an
example, consider the original geometry of the OSHE in which the flux of
$X$-polarized particles having a pseudospin $\mathbf{S}_0=S_x
\mathbf{u}_x$ and a wavevector $\mathbf{k}=k\mathbf{u}_x$ hits an
impurity that redistributes the particles over the elastic circle.
One can easily see from Eqs.(\ref{SEq},\ref{HLT}) that the precession
amplitude of $\mathbf{S}$ is maximal when
$\mathbf{S}\bot\mathbf{H}_{lt}$, namely in diagonal directions, while there is no precession at all in $x$ and $y$ directions where the
vectors $\mathbf{S}$ and $\mathbf{H}_{lt}$ are (anti)parallel.
Consequently, the polarization of the emission becomes strongly
dependent on the scattering angle $\phi$. This leads to the appearance
of alternating centro-symmetric circularly polarized domains in
the four quarters of the $(x,y)$ plane (diagonal directions) in both direct and reciprocal spaces
\cite{OSHE} (equivalent for radial fluxes) [see Figs.\ref{fig1}(a) and \ref{fig3}(b)]. The
formation of these domains is a direct consequence of the onset
of spin currents\cite{OSHENPhys} that are crucial in the context of
designing the future spinoptronic devices.

Instead of using disorder scattering, it is possible to use a
pump spot focused in real space to excite the polariton
dispersion on a ring\cite{Langbein,AOSHE,Sunflower,NOSHE}. The
requirement is that the quasi-resonant injection laser, positioned
at normal incidence ($\mathbf{k}_p=\mathbf{0}$), is blue detuned
from the bottom of the polariton branch. The resulting polaritons
are then propagating radially outward from the spot with a kinetic
energy $E_k=\hbar\omega_p$, where $\omega_p$ is the laser frequency [see
Fig.\ref{fig1}(a,b)]. This setup allows to obtain a homogeneous and
isotropic distribution together with the possibility of
investigating nonlinear regimes where disorder is screened.

The asymptotic analytical solution of the Schr\"{o}dinger equation
for a stationary distribution of the polariton spinor field
$\boldsymbol{\psi}(r,\phi)=(\psi_+,\psi_-)^T$ can be obtained for the case of a
Dirac delta source\cite{NOSHE} at $r=0$:
\begin{eqnarray}
\label{WFTETM1}
{\psi _ + } &=& \sqrt {\frac{{2{N_0}}}{{\pi {k_0}r}}} {e^{ - i\phi }}\left[ {\cos \phi{e^{i{k_l}r}} + i\sin \phi{e^{i{k_t}r}}} \right]{e^{ - r/{r_0}}}\\
\label{WFTETM2}
{\psi _ - } &=& \sqrt {\frac{{2{N_0}}}{{\pi {k_0}r}}} {e^{ + i\phi }}\left[ {\cos \phi{e^{i{k_l}r}} - i\sin \phi{e^{i{k_t}r}}} \right]{e^{ - r/{r_0}}}
\end{eqnarray}
Here $N_0$ is the population imposed by the source depending on the intensity of the pump. ${r_0}=\hbar k_0\tau /{m^*}$
is a mean decay length where $\tau$, $m^*$ and $k_0=(k_l+k_t)/2$ are
the polaritons lifetime, effective mass and mean excitation
wavevector respectively with $k_{l,t}=\sqrt{2 m_{l,t}
\omega_P/\hbar}$ corresponding to the TM ($l$) and TE ($t$) waves
respectively. From Eqs.(\ref{WFTETM1},\ref{WFTETM2}) we immediately
obtain the corresponding distribution of circular polarization
degree of the polariton emission $\rho_c=(n_+-n_-)/(n_++n_-)$ where
$n_{\pm}=|\psi_\pm|^2$:
\begin{equation}\label{rhoc}
\rho_c\left( r,\phi\right)=\sin \left[ {\left({{k_l}-{k_t}}\right)r}\right]\sin \left[ 2\phi  \right]
\end{equation}
revealing the alternating polarization domains. We see that $\rho_c$ is periodic function of both the radial and the angular
coordinates. The corresponding radial frequency
$\nu_r=({k_l}-{k_t})/2\pi$ is associated with the strength of the TE-TM
spitting $\Delta_{lt}$ while the azimuthal one $\nu_\phi=\pi$ is
imposed by the symmetry of the effective magnetic field
$\mathbf{H}_{lt}$. Interestingly, each circular polarization extremum
is associated with a phase dislocation while the total density $n_++n_-$ remains smooth. This is characteristic of a skyrmion
as found in Ref.\onlinecite{NOSHE}.

\section{Polarization inversion}\label{SecII}
The OSHE is a linear effect which does not involve
polariton-polariton scattering. However, taking into account these nonlinear processes can bring qualitative changes to
the related polarization textures and spin currents patterns. We
have recently shown that in a regime where the interactions are
dominant, the polarization currents become strongly focused and
the skyrmions turn into half-solitons\cite{NOSHE}. This effect is due
to the spin-anisotropy of polaritons self-interactions: The
polaritons having the same spin projection interact much more
efficiently than polaritons with opposite
spins\cite{Ciuti,Combescot}, the latter process being of the second order. Moreover, the corresponding matrix
elements can have opposite signs. This feature is responsible for
number of spin dependent nonlinear effects such as polarization
multistability\cite{Multistability}, self-induced Larmor
precession\cite{SILP}, linear polarization build up in polariton
condensates\cite{BECPolaritons} and the inversion of linear
polarization in polariton parametric scattering\cite{Renucci}. The
latter effect is central for the purposes of the present paper as we
will see now.

Due to the their hybrid nature, the shape of the lower polariton
branch is strongly non-parabolic\cite{Microcavities}, which leads
to the appearance of the so-called magic angle close to the inflexion point characterized by
the wavenumber $k_p$. Two identical polaritons
with $\mathbf{k}=\mathbf{k}_p$ can scatter to singlet states with
$\mathbf{k}_s=0$ and $\mathbf{k}_i=2\mathbf{k}_p$ conserving both
the momentum and the energy:
\begin{equation}\label{Conservation}
E\left( {2{\mathbf{k}_p}} \right) = 2E\left( {{\mathbf{k}_p}} \right)
\end{equation}
Therefore, under quasi-resonant pumping at the magic angle, the system is driven into the so-called optical parametric oscillator
regime\cite{OPO} (OPO). We note that with increasing pump power, the
injected mode becomes more and more blueshifted due to polariton-
polariton interactions and therefore the final state selection can
be power dependent\cite{WhittakerPRB}. In polarization resolved OPO
experiments under linearly polarized excitation, the final states
where found to be cross polarized with respect to the
pump\cite{Renucci,PolInv1D} which was theoretically explained using
semi-classical spin-dependent kinetic equations containing the terms
of spin- anisotropic polariton- polariton scattering
\cite{SILP,ShelykhKin,ReviewSpin}. Later on, it was shown that the
effect of polarization inversion is universal and can be
experimentally observed in other pump geometries, e.g. in two pump
horizontal parametric scattering\cite{PolInv}. Combined with
the pseudospin rotation provided by effective TE-TM field, the effect of
parametric polarization inversion can lead to dramatic changes in
the pattern of spin currents as we
will see in the next section.

\begin{figure}[ht]
\includegraphics[width=0.95\linewidth]{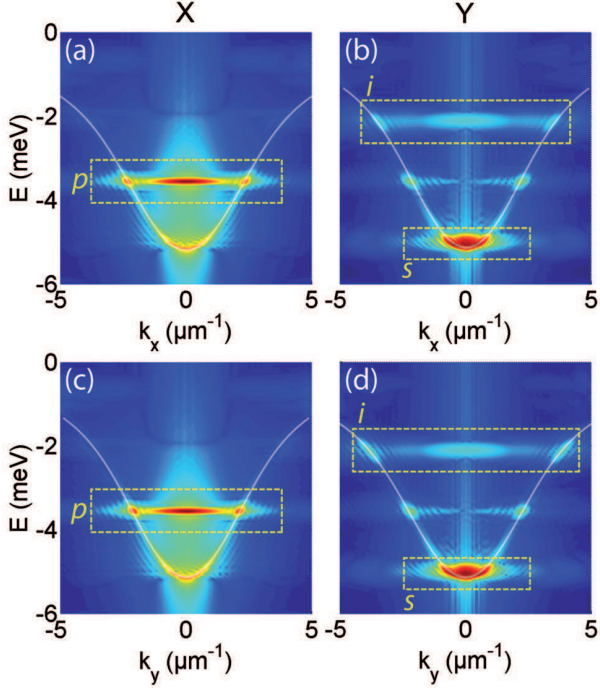}
\caption{(Color online) Time integrated (over 100 ps) $X$ (a,c) and
$Y$ (b,d) dispersions along the $k_x$ (a,b) and $k_y$ (c,d)
directions. The white lines are fits using Eqs.(\ref{Displt}) and
the dashed yellow frames highlight the signal ($s$), pump ($p$) and
idler ($i$) states [see Fig.\ref{fig1}(b)].} \label{fig2}
\end{figure}

\section{The model}\label{SecIII}
We consider the disorder-free microcavity
pumped by a laser spot strongly localized in the real space.
Differently from the situation considered in
Ref.\onlinecite{Langbein}, we take into account nonlinearities
provided by spin-anisotropic polariton-polariton interactions and
focus on the regime where the system is driven to OPO.

In the absence of the TE-TM splitting, the polaritons eigen modes
are degenerated and fully isotropic in the 2D space
$E_l(k_r)=E_t(k_r)$. Therefore using the pumping scheme described in
Sec.\ref{SecI} [Fig.\ref{fig1}(a)] while carefully selecting the pump energy
$\hbar\omega_p$, it should actually be possible to reach the
stationary OPO condition over the whole elastic circle (ring OPO)
under $cw$ excitation. This would give rise to a trichromatic
nonlinear polariton cloud expanding radially from the localized
excitation spot. However, when the energy splitting is taken into
account, as it should be, the TE and TM modes gain slightly different effective
masses $m_\phi^t$ and $m_\phi^l$ and on the linear polarization basis, the
dispersion branches $E_X(\mathbf{k})$ and $E_Y(\mathbf{k})$ become
anisotropic. This is immediately seen from the basis transformation
\begin{equation}\label{xyTETM}
\left( \begin{array}{l}
{E_X}\left( {{k_r},{k_\phi }} \right)\\
{E_Y}\left( {{k_r},{k_\phi }} \right)
\end{array} \right) = \left( {\begin{array}{*{20}{c}}
{\cos \left( {{k_\phi }} \right)}&{\sin \left( {{k_\phi }} \right)}\\
{-\sin \left( {{k_\phi }} \right)}&{\cos \left( {{k_\phi }} \right)}
\end{array}} \right)\left( \begin{array}{l}
{E_l}\left( {{k_r}} \right)\\
{E_t}\left( {{k_r}} \right)
\end{array} \right)
\end{equation}
It means that exciting the system with a linearly
polarized beam, induces a dependence of the OPO condition on
azimuthal angle $k_\phi$. To overcome this angular anisotropy we propose to
use a \emph{pulsed} excitation providing a sufficient energy broadening to encompass the magic point for any azimuthal directions.
We model the quasi-resonant polariton injection with a set of
spin-dependent and driven/dissipative equations for the photonic
$\phi(\mathbf{r},t)$ and excitonic $\chi(\mathbf{r},t)$ fields
coupled via the strong light matter interaction characterized by
the Rabi splitting $\Omega_R=10$ meV:
\begin{eqnarray}
\label{PhiDyn}
\nonumber i\hbar\frac{{\partial {\phi _ \pm }}}{{\partial t}} = &-&\frac{{{\hbar ^2}}}{{2{m_\phi }}}\Delta {\phi _ \pm } + {\Omega_R}{\chi _ \pm } + \beta {\left( {\frac{\partial }{{\partial x}} \mp i\frac{\partial }{{\partial y}}} \right)^2}{\phi _ \mp }\\
 &+& {P_ \pm }{e^{ - i{\omega _p}t}} - \frac{{i\hbar }}{{2{\tau _\phi }}}{\phi _ \pm }\\
\label{ChiDyn}
\nonumber i\hbar\frac{{\partial {\chi _ \pm }}}{{\partial t}} = &-&\frac{{{\hbar ^2}}}{{2{m_\chi }}}\Delta {\chi _ \pm } + {\Omega_R}{\phi _ \pm }\\
 &+& \left( {{\alpha _1}{{\left| {{\chi _ \pm }} \right|}^2} + {\alpha _2}{{\left| {{\chi _ \mp }} \right|}^2}} \right){\chi _ \pm } - \frac{{i\hbar }}{{2{\tau _\chi }}}\chi_{\pm}
\end{eqnarray}
Here $\tau_\chi=400$ ps and $\tau_\phi=20$ ps are the lifetimes of
excitons and photons respectively. The functions
${P}_+(\mathbf{r},t)={P}_-(\mathbf{r},t)$, corresponding to the $X$-
linear polarized pump spot, are 2 $\mu$m large and 20
ps long spatio-temporal Gaussians. The energy of the pump
$\hbar\omega_p$ is blue detuned by an energy $\delta$ (defined below) from the
bottom of the lower polariton branch. $m_\chi=0.4 m_0$, $m_\phi=5\times10^{-5}m_0$ are the
effective masses of the excitons and cavity photons respectively
($m_0$ is the electron mass).

The constant $\beta=\hbar^2/4(1/m_{\phi}^{l}-1/m_{\phi}^{t})$
defines the magnitude of the photonic TE-TM splitting with
$m_{\phi}^{t}=m_\phi$ and $m_{\phi}^{l}=0.95m_\phi$ being the
effective masses of the TE and TM polarized photonic modes. The
corresponding terms give rise to the in-plane effective magnetic
field $\mathbf{H}_{LT}(\mathbf{k})$. The constants
$\alpha_1=6\times10^{-3}$ meV$\cdot\mu$m$^2$ and
$\alpha_2=-0.2\alpha_1$ define the strength of the interaction
between polaritons of the same and opposite circular polarizations respectively.
All the values of the parameters we consider are typical for GaAs
based microcavities.

To find the OPO condition we need to know the bare dispersion
relations of the linearly polarized modes. We first find the
dispersion relations for the TE and TM polariton modes diagonalizing
the $2\times2$ Hamiltonian corresponding to the exciton-photon coupling
\begin{equation}\label{TETMMat}
{M_{l,t}} = \left( {\begin{array}{*{20}{c}}
{E_\phi ^{l,t}\left( {{k_r}} \right)}&{{\Omega _R}}\\
{{\Omega _R}}&{E_\chi \left( {{k_r}} \right)}
\end{array}} \right)
\end{equation}
where $E_\phi ^{l,t}=\hbar^2 k_r^2/2m_\phi^{l,t}$ are photonic dispersions
for TE and TM polarized modes, and $E_\chi =\hbar^2 k_r^2/2m_\chi$
is the excitonic dispersion, for which we supposed the longitudinal-
transverse splitting to be negligibly small. The resulting
polariton dispersions are given by the well known expressions
\begin{equation}\label{Displt}
{E_{l,t}}\left( {{k_r}} \right) = \frac{E_\phi ^{l,t} + E_\chi}{2} - \frac{1}{2}\sqrt {{{\left( {E_\phi ^{l,t} - E_\chi} \right)}^2} + 4\Omega _R^2}
\end{equation}
which reveal the strong nonparabolicity in the vicinity of the
inflexion point. The transformation to the basis of $X$ and $Y$
polarized states can then be done using Eq.(\ref{xyTETM}).

The energy of the angle dependent OPO conditions is found from the
solution of Eq.(\ref{Conservation}). For the previously defined
parameters, we obtain that depending on the angle $k_\phi$ the magic
point lies in the interval of energies ${\Delta E_{OPO}} =
\left[ { - 3.88, - 3.78} \right]$ meV and the corresponding interval
of wavevectors ${\Delta k_{OPO}} = \left[ {1.49,1.58}
\right]\mu$m$^{-1}$. In the numerical simulations we then impose a
detuning $\delta=1.17$ meV.
We consider an intermediate pump power slightly above the OPO
threshold in order to remain in a regime where the interactions are
not dominant and to avoid the onset of the strong spin
focusing\cite{NOSHE}. Finally, we note that
Eqs.(\ref{PhiDyn},\ref{ChiDyn}) don't allow for spontaneous
scattering by themselves and to trigger the OPO process, we have
added to them an additional term corresponding to a weak background
gaussian-correlated noise. A weak $Y$ polarized probe at $k=0$ could be used as well\cite{TOPO}.

\begin{figure}[ht]
\includegraphics[width=0.95\linewidth]{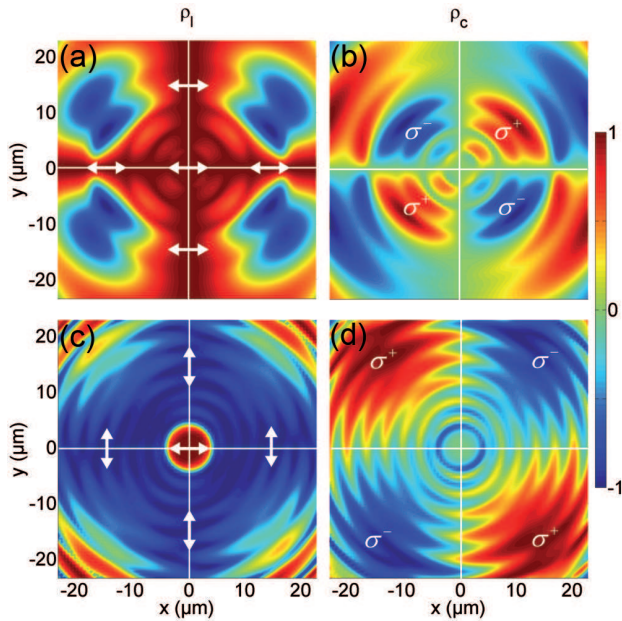}
\caption{(Color online) Degrees of linear $\rho_l$ (a,c) and circular $\rho_c$ (b,d)
polarization of the photonic polariton components shown before (a,b) ($t=15$ ps)
and after (c,d) ($t=45$ ps) the polarization inversion. The double
arrows show the polarization directions that are rotated by $\pi/2$
between (a) and (c) in cross directions. In the diagonal directions
the circularly polarized domains are inverted between (b) and (d)
highlighting the spin current inversion.} \label{fig3}
\end{figure}

\section{Results and discussion}
The results are presented in Figs.\ref{fig1}(c), \ref{fig2} and
\ref{fig3}. The panel (c) of Fig.\ref{fig1} shows the space integrated photonic density
of the $X$ (solid blue line), $Y$ (solid pink line) polarized polariton and the sum of the two (black line) versus time revealing the
$X$-polarized pulsed excitation and the polarization inversion
produced by the OPO. Indeed, initially the polariton emission is mainly $X$-polarized until the OPO stimulation becomes strong from about 25 ps, when it starts to be dominated by the $Y$ polarization due to the rotation of the pseudospin of the scattered polaritons.

The Fig.\ref{fig2} shows slices of the $X$ and $Y$ dispersions
[panels (a,c) and (b,d) respectively] along the $k_x$ and $k_y$
direction [panels (a,b) and (c,d)] integrated over 100
ps. These representations capture most of the OPO and inversion features
in a single representation. We clearly see the pump ($p$) state in the $X$-component and
the signal ($s$) and idler ($i$) states appearing in the $Y$ component. Note
that although we pump with $X$-polarized light, the pump state is
slightly visible in the $Y$-component as well. This is simply due to the $X$ to $Y$
polarization conversion provided by TE-TM splitting away from the spot prior the inversion [blue regions in Fig.\ref{fig3}(a)].

The Fig.\ref{fig3} shows the degrees of linear
$\rho_l=\Re(\phi_+\phi_-^*)/(n_{\phi+}+n_{\phi-})$ [panels (a,c)] and circular
$\rho_c=(n_{\phi+}-n_{\phi-})/(n_{\phi+}+n_{\phi-})$ [panels (b,d)] polarization of
the photonic component (the quantity measured experimentally) at
$t=15$ ps and $t=45$ ps respectively (vertical dashed lined in
Fig.\ref{fig1}(c)). Before the polarization inversion (onset of the
OPO) [panels (a,b)] the diagonal polarization domains are those expected for the linear OSHE,
while as soon as the OPO is triggered [panels (c,d)], at the edges of the spot the $X$-polarized pump
polaritons are instantly converted to the $Y$-polarized polaritons
in the signal and idler states. We then have a situation equivalent
to the excitation in $Y$ polarization corresponding to $\mathbf{S}_0=-S_x \mathbf{u}_x$. Consequently, as can be seen
from Eq.(\ref{SEq}) the pseudospin precession is reversed under $\mathbf{S}\rightarrow-\mathbf{S}$ and
the circular polarization domains become inverted ($\sigma_+\leftrightarrow\sigma_-$) which corresponds to an inversion of the spin currents. Importantly, the outgoing nonlinear waves are bichromatic and have two different
wavevectors associated to the signal and idler states, which explains the interferences visible in the panel (d). The pump state is emptied due to the pulsed excitation [see Fig.\ref{fig1}(c)]. The dominant contribution to the measurable photonic component is the signal state since the $k$-dependent photonic fraction $F_\phi(k)$ of polaritons decreases from 0.5 in the signal to less than 0.1
in the idler. It can be easily checked finding the eigenvectors of $M_{l,t}$ and yielding
\begin{eqnarray}
{F_\phi } = \frac{{4\Omega _R^2 - \Delta {E_{l,t}}\left( {\Delta {E_{l,t}} + \sqrt {\Delta E_{l,t}^2 + 4\Omega _R^2} } \right)}}{{2\Delta E_{l,t}^2 + 8\Omega _R^2}}
\end{eqnarray}
where $\Delta E_{l,t}=E_\phi^{l,t}-E_\chi$. When the OPO scattering occurs, most of the emission comes from the $Y$-polarized signal state close to $k=0$, which explains why the polarization domains become
more extended in real space [compare panels (b) and (d)]. Indeed, the $k$-dependent precession becomes slower [see Eqs.(\ref{DeltaLT},\ref{Displt})] in this state as expected for a reduced value of $\Delta_{lt}(k)$.

We should stress that we worked in a regime where the compression of
the polarization domains due to nonlinearities is weak which corresponds to the formation of a skyrmion
lattice in terms of Ref.\onlinecite{NOSHE}. Therefore, we consider
the interval of pump powers intermediate between the linear regime
considered in Ref.\onlinecite{Langbein} and strongly nonlinear
regime considered in Ref.\onlinecite{NOSHE}. Interestingly, during the inversion the phase singularity of each skyrmion is transferred from one component to the other. Additionally, one can expect that a further increase of the pump in the OPO regime would lead to the focusing of the inverted spin currents. This would actually compete with the relatively small group velocity in the signal state that makes the polaritons decay before the non-linear effects, that trigger the spin focusing, start to play a strong role.

\section{Conclusions}
In summary, we have shown that the optical parametric oscillations
can be triggered on a whole elastic circle in planar microcavities
under pulsed excitation. This regime is associated with the onset of
linear polarization inversion leading to the inversion of the spin
currents of the optical spin-Hall effect. Together with the effect
of nonlinear focusing described earlier in Ref.\onlinecite{NOSHE}, we have
proposed two mechanisms allowing to control the spin currents in
semiconductor microcavities which could be crucial for future spinoptronic
applications.

\section{Acknowledgments}
We acknowledge the support of the FP7
ITN Spin-Optronics (237252) and IRSES "POLAPHEN" (246912) project.
I. A. S. benefited from Rannis "Center of Excellence in
polaritonics" and IRSES "SPINMET" project.

\end{document}